\documentclass[11pt]{amsart}
\usepackage{amsfonts}
\usepackage{eurosym}
\usepackage[utf8]{inputenc}
\usepackage[english]{babel}
\usepackage[document]{ragged2e}
\usepackage{ulem}
\usepackage{amsmath}
\usepackage{amssymb}
\usepackage{fontenc}
\usepackage{graphicx}
\usepackage{color}
\usepackage{xcolor}
\usepackage{amsthm}
\usepackage{enumitem}
\usepackage{algorithm}
\usepackage{algorithmic}
\usepackage{hyperref}
\usepackage{shortvrb}
\usepackage{endnotes}
\usepackage{setspace}
\usepackage{makecell}
\usepackage{nicefrac}
\usepackage[foot]{amsaddr}
\usepackage{hyperref}

\setitemize[0]{labelindent=20pt,leftmargin=*,align=left,itemsep=0.2em}

\newcommand{\RNum}[1]{\uppercase\expandafter{\romannumeral #1\relax}}

\makeatletter
\def\@textbottom{\vskip \z@ \@plus 1pt}
 \let\@texttop\relax
\makeatother

\newtheorem{theorem}{Theorem}

\makeatletter
\@namedef{subjclassname@2020}{  \textup{2020} Mathematics Subject Classification}
\makeatother
\email{eduardo.salazar@forctis.io}
\thanks{The authors wish to thank Forctis AG for providing funding to this project.}

\title[Design and implementation of a novel CSPRG]{\textbf{Design and
implementation of a novel cryptographically secure pseudorandom number
generator}}
\author{Juan Di Mauro$^1$}
\address{$^1$Instituto de Ciencias de la Computaci\'on \\
Universidad de Buenos Aires and CONICET, Buenos Aires, Argentina.}
\author{Eduardo Salazar$^2$}
\address{$^2$Forctis AG, Wollerau, Switzerland (Corresponding author)}
\author{Hugo D. Scolnik $^{1,2,3}$}
\address{$^3$Departamento de Computaci\'on\\
Facultad de Ciencias Exactas y Naturales, Universidad de Buenos Aires,
Buenos Aires, Argentina.}
\date{}

\begin{document}

\maketitle

\begin{abstract}
The aim of this paper is to present a new design for a pseudorandom number
generator (PRNG) that is cryptographically secure, passes all of the usual
statistical tests referenced in the literature and hence generates high
quality random sequences, that is compact and easy to implement in practice,
of portable design and offering reasonable execution times. Our procedure
achieves those objectives through the use of a sequence of modular
exponentiations followed by the application of Feistel-like boxes that mix
up bits using a nonlinear function. The results of extensive statistical
tests on sequences of about $2^{40}$ bits in size generated by our algorithm
are also presented.
\end{abstract}

\justify

\section{Introduction}

A \textit{pseudorandom number generator} (PRNG) is a polynomial time
computable function $f$ that maps a short random string $x$ into a long one $%
f(x)$ that appears to be random (patternless) to any external observer. In
other words, the output sequence of a PRNG should be indistinguishable from
a \textit{truly random} sequence for any polynomial-time algorithm. In turn,
a sequence is truly random if it is the realization of a Bernoulli process
with success probability equal to $1/2$. The output of a PRNG can be
therefore seen as a finite sequence of bits, such that each bit has to be
independently generated with equal probability of being a $0$ or a $1$. See,
amongst others, chapter 5 in \cite{menezes}. The term pseudorandom is often
used because such sequences are generated by means of deterministic
algorithms.

The generation of \textit{random sequences} is an essential ingredient for
many scientific applications (for example, computer simulations, statistical
sampling, stochastic optimization and cryptography, only to cite a few).
There is a marked tendency amongst practitioners to focus on the speed at
which random bits can be generated rather than on the true randomness of
those bits. Still, there are many situations where the quality of generated
random numbers has a more direct, and likely irreversible, impact on the
system or processes to which they are applied. One such case are \textit{%
cryptographic systems}, because the quality of such random bits determines
how prone those systems are to a successful attack.

At this stage, the notion of \textit{unpredictability} comes into play,
meaning that a value should be very difficult to guess by an attacker. It
has two sides to it: on the one hand, the knowledge of the first $k$
elements of a sequence should make it infeasible to predict what the $k+1$
element in that sequence would be with probability greater than $1/2$ (known
as forward unpredictability or next-bit test; on the other, it should also
be impossible to determine the seed used by an algorithm from the knowledge
of any sequence of bits generated by it (or backward unpredictability). But,
if all the practitioner has at hand is a deterministic algorithm and digital
hardware, generating a true random sequence becomes a hard problem.%
\cite{TRG}

Perhaps the obvious choice would be to use one (or ideally, several)
statistical test(s) to determine whether a random sequence generated by a
computer algorithm is as random as it could possibly be, given it is
generated by a finite-state machine. Many competing \textit{protocols} (a
structured group of tests, or a \textit{battery} as they are often called)
have been so far designed, that in addition are also freely available. The
suites most frequently referenced in the literature are outlined below,
together with the respective links and in no particular order of preference.

\begin{itemize}
\item \textbf{NIST SP 800-22 Rev. 1a}. \cite{NISTtest}

\item \textbf{TestU01} by P. L'Ecuyer and R. Simard. \cite{TestU01}

\item \textbf{Practically Random}, popularly known as PractRand. \cite%
{PractRand}

\item \textbf{Random Bit Generators Tester} or RaBiGeTe. \cite{RaBiGeTe}

\item \textbf{Diehard} by George Marsaglia. \cite{DieHardM}

\item \textbf{DieHarder} by Robert Brown, a cleaned up and enhanced version
of Marsaglia's Diehard. \cite{DieHardB}
\end{itemize}

\noindent Unfortunately, none of those tests can really \textit{prove} that
a sequence is \textit{truly random}. Why? Because it is rather simple to
generate bits that could pass statistical tests for randomness and yet are
also perfectly predictable. This can be seen through a trivial, almost
absurd example: a string made of $N$ repetitions of the number sequence ${%
0123456789}$. Intuitively, in a random sequence the digits $0$ to $9$ should
appear with approximately equal frequency; in other words, the underlying
distribution of digits should be approximately $\mathcal{U}\left( 0,9\right) 
$. It is easy to see this condition holds (as a sequence generated in this
way has $N$ 0's, $N$ 1's,\ldots , $N$ 9's) yet it can be also easily argued
that such sequence is perfectly predictable. This happens because the
generated sequence uses a \textit{low entropy} source; or to be more
precise, this particular statistical check for randomness is completely
blind to the source of entropy used to generate the sequence.

The reader might now be tempted to ask the following question: how can it be
guaranteed that a sequence is both patternless and unpredictable? An answer
was proposed by A. Kolmogorov, G. Chaitin and R. Solomonoff, who
independently reasoned that any sequence computed in a finite state machine
cannot be truly random in the sense of the theoretical definitions of
randomness [as explained, for instance, in \cite{chaitin}]. On that basis, a
definition\ (now widely known as \textit{Kolmogoroff complexity}) has been
proposed: a sequence (a series of numbers, symbols or both) is random if the
smallest algorithm $K(x)$ capable of specifying it to a computer has about
the same number of bits $\left\vert x\right\vert $ of information as the
series $x$ itself. [see \cite{chaitin2}]

Following this definition, sequences should not be considered random when $%
K\left( x\right) \ll x$, that is, when an algorithm $K$ can be described
(written) using\ substantially fewer bits than the sequence it
generates as output. In other words, to be considered as random a sequence
must be \textit{incompressible}. But, once again, there is a catch:
Kolmogorov complexity cannot be computed. Why? Because one can never be
completely sure to have found the shortest computer program capable of
describing a string of length $x$. Despite this practical inconvenience the
principle is still conceptually very relevant.

The preceding discussion simply highlights the importance of understanding
that passing any battery of statistical tests is simply not enough. The fact
is that the majority of generators shown in \cite{pcg} including, amongst
them, the popular Mersenne Twister (the standard, and most widely used\
version, is MT19937, 32-bit) fail some statistical tests of randomness, at
least when examined using the protocols enumerated above, and are unsuitable
for cryptographic applications straight out of the box. Hence Matsumoto and
Nishimura, authors of the Mersenne algorithm, explicitly suggest in their
original paper \cite{mersennetwister} to include hashings as a fool-proof
mechanism to make their generator cryptographically secure. It should also
not be too difficult to note that a hashing routine can be easily added as
the final step to any PRNG, not just to the Mersenne Twister. Hashing
functions might be quite useful for this task, however, in the end they just
serve to hide weaknesses that might be otherwise present.

Taking all of the above into account, our aim was to develop a PRNG that is
(a)\ statistically sound; and (b)\ relies on the intractability of a
number-theoretic problem to provide cryptographic security. What is known as
a \textit{cryptographically secure pseudorandom number generator}
(abreviated CSPRNG).

\section{Preliminary considerations}

There are many important papers in this specific field, for instance \cite%
{zuckerman} and generalizations seeking to extract bits from different
pseudorandom sources. Several authors have proposed using modular
exponentiations as a mechanism to generate pseudorandom numbers, like the
Blum-Micali algorithm \cite{blummicali} which extracts one bit per
iteration. Despite being very important theoretical contributions, all share
a major drawback: they are very slow for practical applications, even if one
considers increasing the number of output bits as in \cite{patel}.

Bearing all of this in mind, we found \cite{beale} of particular interest
because it presents several remarkable (and rather easy to implement) ideas.
Our approach explored the possibilities of using consecutive modular
exponentiations, both in simple and multiple precision. Soon it became clear
that by simply using only modular exponentiations some undesirable
statistical problems persisted. Recalling the Feistel cipher that was used
in the DES (Data Encryption Standard), although in DES with a very poor
nonlinear function [refer to \cite{menezes} for a detailed discussion], we
designed a generalized version based on a different nonlinear function to
generate the final output after each iteration (see \texttt{Algorithm \ref{algo1}}
in the next section).

\section{The proposed algorithm}

We now provide some preliminary comments regarding our design. For any given
seed, $v=1,\ldots ,V$ initial \textit{safe primes} $p_{v}\in \mathcal{P}$
are either retrieved from a pre-calculated table (storing, for example, one
million safe primes) or generated as needed. In practical applications, our
advice is for $V\geq 6$. Recall that safe primes are primes $p>2$ such that $%
(p-1)/2$ is also a prime.

When using a table of safe primes, indices refer to a subset of that table.
If the safe primes are otherwise generated with each run, we propose to
compute them within a certain interval, not necessarily arbitrary. Those
primes change throughout the number generation process.

\newpage
If required, the generators for $\mathbb{Z}_{p}^{\ast }$ can be fixed or
calculated according to the following results [see \cite{boris}].

\begin{itemize}
\item If $p\geq 7$ is a safe prime, then $g=p-\lfloor\sqrt(p)\rfloor^2$ is a
generator.

\item If $p\geq 7$ is a safe prime, for every integer $z$ that satisfies the
inequalities $2\leq z\leq p-2$, $g=(p-z^2)\mod p$ is a generator.
\end{itemize}

\vspace*{2mm}
\begin{algorithm}
\caption{$PRNG(n,x_0,\{i_0,\ldots,i_{s-1}\},s)$} 
\label{algo1}
\begin{flushleft}
\textbf{REQUIRE} (as parameters)$\>$:\\
\vspace*{1.5mm}
$M_R$ = max number of elements generated for a subset of indexes.\\
$nrounds$ = the number of Feistel-like rounds.\\
$q$ = a big safe prime (in practice, \textit{big} implies a prime $\geq$ 1024 bits).\\
$g$ = a generator of $\mathbb{Z}_q^*$.\\
$\mathcal{P}$ = a set of safe primes of size $k$ in bits (where $k\geq $32).\\
$n$ = the number of random elements to generate.\\
$x_0$ = an initial element.\\
$\{i_0,\ldots,i_{s-1}\}\subset\{0,\ldots,|\mathcal{P}|-1\}$ = a set of indexes.\\

\end{flushleft}
\begin{algorithmic}[1]
 \STATE Let $t\leftarrow \log_2{q}-k$.\label{algo1:0}
  \STATE $p$ a safe prime of length in bits greater than $k$, $a$ a generator of $\mathbb{Z}^*_p$ \label{algo1:1}
  \STATE $e_1\leftarrow 17$, $e_2\leftarrow 9$ \label{algo1:2}
  \STATE $i\leftarrow 0$\label{algo1:3}
  \WHILE{$i<n$}\label{algo1:4}
    \IF{$i>0$ and $i\mod M_R\equiv 0$}\label{algo1:5}
        \STATE Obtain randomly $s$ indexes $\{i_0,\ldots,i_{s-1}\}\subset\{0,\ldots,|\mathcal{P}|-1\}$\label{algo1:6}.
    \ENDIF\label{algo1:7}
    \STATE $w_1\leftarrow aw_1\mod p$.\label{algo1:8}
    \STATE $x_0\leftarrow x_0+w_1\mod p_{i_0}$\label{algo1:9}
    \FOR{$j=1,\ldots,s-2$}\label{algo1:10}
    \STATE $x_j\leftarrow x_{j-1}^{e_1}\mod p_{i_j}$\label{algo1:11}
    \ENDFOR
    \STATE $x\leftarrow x_{s-3}|x_{s-2}$.\COMMENT{The bar "$|$" is the bit concatenation operator}
    \STATE Discard the first bit of $g^{x}\mod q$ and let
    $z$ be the $t$ least significant bits that remains\label{algo1:14}.
    \STATE Let $z_1,\ldots,z_m$ be integers of length $k$ such that $z=z_1|\ldots|z_m$.  \label{algo1:15}
    \STATE $x_0\leftarrow f(z_1,p_{i_{s-1}})$\label{algo1:16}
    \FOR{$j=2,\ldots,m$}
        \STATE $y_i\leftarrow f(z_j,p_{i_{s-1}})$ \label{algo1:17}
        \STATE $i\leftarrow i+1$.
    \ENDFOR
  \ENDWHILE\label{algo1:24}
  \RETURN $\{y_i\}_{i=1}^n$ \label{algo1:25}
 \end{algorithmic}
\end{algorithm}

\begin{algorithm}
\caption{Feistel-like box iterations $f(x,n)$}
\label{algo2}
 \begin{algorithmic}[1]
 \STATE $r_0\leftarrow x[k/2+1,k]$ \COMMENT{$x[i,k]$ are the bits $i,i+1,\ldots,k$ of $x$}\label{algo3:1}
    \STATE $l_0\leftarrow x[0,k/2]$\label{algo3:2}
    \FOR{$i=1,\ldots,nrounds$}\label{algo3:3}
        \STATE $l_1\leftarrow r_0$\label{algo3:4}
        \STATE $w_2\leftarrow aw_2\mod p$\label{algo3:5}
        \STATE $x\leftarrow (w_2+(r_0\oplus l_0))^{e_2}\mod n$\label{algo3:6}
        \STATE $x\leftarrow (l_0\oplus x)$\label{algo3:7}
        \STATE $r_1\leftarrow x[0,k/2]$\label{algo3:8}
        \STATE $x\leftarrow l_1|r_1$\label{algo3:9}
        \STATE $l_0\leftarrow l_1$, $r_0\leftarrow r_1$\label{algo3:10}
    \ENDFOR\label{algo3:11}
    \RETURN $x$
 \end{algorithmic}
\end{algorithm}

In line \texttt{\ref{algo1:14}} of \texttt{Algorithm \ref{algo1}}
note that $t$ has to be set bearing in mind the number of hard bits of the
one-way function in a Blum-Micali scheme. Here, as the primes $%
p_{i_{0}},\ldots ,p_{i_{s-1}}$ are intended to be small and (in accordance
with the discussion in the next section) we only regard $g^{x}\mod(q)$ as
the one-way function and not the composition of the $s$ exponentiations
beginning in line \texttt{\ref{algo1:10}}. So the same
considerations about hard bits in \cite{patel} apply here and we can take $%
t=\log {q}-k$. Also, $w_{1}$ in \texttt{Algorithm \ref{algo1}} and $w_{2}$
in \texttt{Algorithm \ref{algo2}} are two integers such that $0<w_{1}$, $%
w_{2}<p$.

Our PRNG can therefore be seen as an application from $\{0,1\}^{\ell}$ to $%
\{0,1\}^L$ that maps a bit string $w$ composed by a seed $0\leq x\leq2^{32}$
and a set of indexes $i_0,\ldots,i_{s-1}$ with $0\leq i_j \leq |\mathcal{P}%
|-1$ to a bit string $v$ of length $L$. Hence, if $|\mathcal{P}|=2^r$ then $%
|w|\leq\ell=k+sr$ and $L=k(m-1)M_R$ (we choose $M_R=2^{11}$ and note that $%
z_1$ in line \texttt{\ref{algo1:16}} is not used to produce the
output).

Concerning the change of base, the simplest option is to use an iteration
counter. There are, of course, more sophisticated alternatives, such as
chaotic functions, non-linear mappings or oscillators (even if based on some
form of deterministic input). In practice, we have nevertheless found that
an iteration counter is enough to deliver robust sequences.
 
\noindent \texttt{Important remark} $-$ The word \textit{randomly} on line
\texttt{\ref{algo1:6}} in \texttt{Algorithm \ref{algo1}} can be made
precise by replacing it with the following:

\begin{itemize}
\item Do the steps from \texttt{\ref{algo1:8}} to \texttt{\ref{algo1:15}}.

\item Let $i_j\leftarrow f(z_j,p_{i_{s-1}})$ for $j=0,\ldots,s-1$.
\end{itemize}

\noindent in order to allow the $s$ indexes to be obtained by the same
process that generates the random values.

Finally, the exponents $e_1$ and $e_2$ are suitably chosen to make the
non-linear function in line \texttt{\ref{algo1:9}} of \texttt{Algorithm \ref{algo1}} different
to the non-linear function in \texttt{Algorithm \ref{algo2}}. The binary representation
of the exponents have only two $1$'s allowing for a fast evaluation of the modular
exponentiation.

\section{Security of the PRNG}

Let $\mathcal{P}$ be a set of safe primes, as described before. The \textit{%
seed space} is given by the size of the primes and the initial value $x_{0}$.
If $2^{r}=|\mathcal{P}|$ and the elements of $\mathcal{P}$ are of $k$ bits
of size, then the key length is equal to $2^{k+sr}$. In order to achieve a
complexity equivalent to a key length of $128$ bits, for $k=32$ and since $%
r=21 $ (because there are around $2^{21}$safe primes of $32$ bits) the
number of primes and indexes must be at least $s=6$.

Shifting our focus on \textit{cryptographic security} and for the sake of
completeness, it is useful to recall at this stage the \textit{Discrete
Logarithm Problem} (hereafter DLP).

\noindent \texttt{Discrete Logarithm Problem} $-$ Given $g,z,n\in \mathbb{Z}$
find $x$ such that $g^{x}\equiv z\mod q$ if such an $x$ exists.

It is important to note here that the primes $p_{i_0},\ldots,p_{i_{s-1}} $
remain hidden to an external observer. Even assuming that an attacker is
able to extract them by means of cryptoanalysis, line
\texttt{\ref{algo1:14}} of \texttt{Algorithm \ref{algo1}} generates a series of
output bits that are simultaneously hard with respect to the modular
exponentiation $g^x\mod q$. Consequently, predicting those bits is \textit{%
at least as hard} as solving the DLP for those numbers.

The procedures that attempt to solve the DLP problem can be classified into
different categories. There are algorithms for general groups without
special characteristics like the Baby-step Giant-step algorithm [according
to \cite{od} due originally to D. Shanks]; Pollard's rho factoring algorithm 
\cite{pollard}; methods for finite groups whose orders have no large prime
factors as Pohlig-Hellman's \cite{pohlig}; and the subexponential algorithms
like Adleman's index calculus, that use the Chinese Remainder Theorem \cite%
{adleman} and Gordon's Number Field Sieve.

We focus on the case when $n$ is a prime number, and all those algorithms
share the fact that complexity increases if $n-1$ has large factors. This
lends support to the use of safe primes, because $p-1$ has a factor of order 
$p$. To have any chance at solving the DLP problem, an attacker should at a
minimum know both the prime and the generator being used.

\subsection{Security bounds}

Let us denote by RG the PRG given in \cite{Gennaro} but rewrite the
procedure offered to generate random bits. Let $p$ be a strong prime of $n$
bits and $g$ a generator of the group $\mathbb{Z}_p^*$. Let $t=n-c=2^l$
where $c$ is a quantity that grows faster than $\log_2 n$; that is, $%
c=\omega(\log_2 n)$ and as usual we take $c=128$ for a $1024$ bit prime $p$.

\begin{algorithm}
\caption{RG generator}
 \begin{algorithmic}
  \REQUIRE  $x_0\in \mathbb{Z}_{p-1}$ as seed, $n>0$ (number of random words to output).
   \FOR{$i=1,\ldots,n$}
   \STATE Let $\hat{g}\leftarrow g^t \mod p$
   \STATE Let $b_1$ be the first bit of $x_{i-1}$
   \STATE Let $x_{i}=\hat{g}^{x_{i-1} div\ t}g^{b_1}$
   \STATE Let $r_i$ be the $t-1$ least significant bits of $x_{i}$ after discharging the first (left) bit.
   \ENDFOR
   \STATE Output $r_1,\ldots,r_n$
 \end{algorithmic}
\end{algorithm}

Our generator uses RG as a subroutine, as can be easily seen in line
\texttt{\ref{algo1:14}} of \texttt{Algorithm \ref{algo1}}. We will prove here that,
actually, this RG generator provides an upper bound of the security of our PRG.
The proof is a mild reduction strategy that builds an adversary for RG from an
adversary to our generator.

In passing, note that some minor calculations are needed to map line
\texttt{\ref{algo1:14}} of \texttt{Algorithm \ref{algo1}} to the RG. Those calculations are
polynomially bounded in the length of $n$, meaning that the map between this line in our
generator and the RG is a polynomial map. Therefore, the operations required for its
computation (using RG as a subroutine) are bounded by $p(\log n)$ for some polynomial $p$.
Details are given in the Appendix.

We recall here the notion of a secure pseudo-random function for
convenience. A pseudo-random function (or PRF for short) is a family of
mappings from some space $\mathcal{X}$ to another set $\mathcal{Y}$. The
members of the family can be indexed by a set of keys $\mathcal{K}$, so,
given $k\in \mathcal{K}$ the elements $f_k$ of the family can be written as $%
F(k,\cdot)$, and the entire family can be defined as
\vspace*{2mm} 
\begin{equation}
PRF=\{F(k,\cdot):\mathcal{X} \rightarrow \mathcal{Y} \mid k\in \mathcal{K}\}
\end{equation}

\noindent (note that $\mathcal{X}$ could be equal to $\mathcal{Y}$ although
that does not imply that both functions are permutations). Usually, the set
of all functions from $\mathcal{X}$ to $\mathcal{Y}$ is written as $Funcs[%
\mathcal{X},\mathcal{Y}]$ so a PRF is a (special) subset of $Funcs[\mathcal{X%
},\mathcal{Y}]$.

In turn, a PRF is deemed secure if there is no adversary that can tell the
difference between $F(k,\cdot)$ and any random function in $Funcs[\mathcal{X}%
,\mathcal{Y}] $ provided, of course, that $k\in \mathcal{K}$ is randomly
chosen. We will not provide a more formal definition, to keep with the pace
of the presentation; the interested reader can look at \cite{Boneh},
Definition 4.1.1, which is the one that we will use in this paper.

The advantage of an adversary $\mathcal{A}$ when differentiating between the
families $F$ and $Funcs[\mathcal{X},\mathcal{Y}]$ after evaluating $Q$
elements of its choice is written as $PRFadv[\mathcal{A},F]$. We say that
the PRF $F$ is secure if that advantage is negligible.

By strengthening the restrictions under which and adversary and its
challenger interact, we can relax the conditions over the PRF. Hence, if an
adversary is only allowed to query random points in the domain of the PRF
family then the PRF is weakly-secure when such advantage, written as $%
PRFadv_{weak}[\mathcal{A},F]$, is negligible.

An important point to note is that a secure PRF is also a weakly-secure PRF,
however the reverse implication does not generally hold. In practice,
finding a weakly-secure PRF is supposed to be far easier than finding a
secure PRF.

Bearing that in mind, the family of functions that \texttt{Algorithm \ref{algo2}} represents,
denoted by
\vspace*{2mm} 
\begin{equation}
\{f(p,\cdot ):\{0,1\}^{k}\rightarrow \{0,1\}^{k}\mid p\mathnormal{\ }\text{%
\textnormal{a safe prime}}\leq 2^{k}\}
\end{equation}

\noindent is a PRF. Moreover, our PRG design does not rely on this
particular function but only on a property of this function, which we use in
step \texttt{17} of \texttt{Algorithm \ref{algo1}}. At this point
it is possible to state the following:

\begin{theorem}
If the RG generator is secure and the family of functions $f$ is secure,
then the PRG called $G$ given by \ref{algo1} is secure.
\end{theorem}

\noindent \texttt{Important remark} $-$ This Theorem relies on the fact that
for every polynomial-time adversary $\mathcal{A}$ to $G$ there exists an
adversary $\mathcal{B}$ to the RG and an adversary $\mathcal{B}^{\prime }$
to the PRF $f$ which are elementary wrappers around $\mathcal{A}$, such that 
\vspace*{2mm} 
\begin{equation*}
PRGadv[\mathcal{A},G]\leq PRGadv[\mathcal{B},RG]+PRFadv_{weak}[\mathcal{B}%
^{\prime },f]
\end{equation*}

\noindent We proceed to build our proof using sequence of games that
describe the interaction between a challenger $Ch$ and the adversary $%
\mathcal{A}$.

\begin{proof}
\noindent Let $Game\ 0$ be defined by the following sequence of steps:

\begin{itemize}
\item The challenger receives $x_0$ from $\mathcal{A}$.

\item The challenger runs one round of the generator (lines \texttt{9} to
\texttt{21} of \texttt{Algorithm \ref{algo1}}) with $y_1,\ldots,y_m$ as the result.

\item The challenger sends $y_1,\ldots,y_m$ to $\mathcal{A}$.

\item $\mathcal{A}$ outputs one bit $b$.
\end{itemize}

\noindent In turn, $Game\ 1$ proceeds as follows:

\begin{itemize}
\item The challenger receives $x_0$ from $\mathcal{A}$.

\item The challenger sets $r_1,\ldots,r_m$ as random elements of $\{0,1\}^k$.

\item The challenger runs one round of the generator (lines \texttt{9} to
\texttt{21} of \texttt{Algorithm \ref{algo1}}). For $i=1,\ldots,m$ it replaces $z_i$ with $r_i$ in
line \texttt{16}. Let $y_1,\ldots,y_m$ be the result.

\item The challenger sends $y_1,\ldots,y_m$ to $\mathcal{A}$.

\item $\mathcal{A}$ outputs one bit $b$.
\end{itemize}

\noindent Finally, $Game\ 2$ evolves according to the following sequence:

\begin{itemize}
\item The challenger receives $x_0$ from $\mathcal{A}$.

\item The challenger sets $r_1,\ldots,r_m$ as random elements of $\{0,1\}^k$.

\item The challenger runs one round of the generator (lines \texttt{9} to
\texttt{21} of \texttt{Algorithm \ref{algo1}}). Let $y_1,\ldots,y_m$ be the result.

\item The challenger sends $r_1,\ldots,r_m$ to $\mathcal{A}$ (it replaces $%
y_{i}$ with $r_i$).

\item $\mathcal{A}$ outputs one bit $b$.
\end{itemize}

\noindent Let $W_i$ be the event where $\mathcal{A}$ answers $1$ in $Game\ i$%
. It becomes clear that
\vspace*{2mm} 
\begin{equation}
PRGadv[\mathcal{A},G]=|Pr[W_0]-Pr[W_2]|
\end{equation}

\noindent and that the above equation can be rewritten as
\vspace*{2mm} 
\begin{equation}
PRGadv[\mathcal{A},G]=|Pr[W_0]-Pr[W_1]+Pr[W_1]-Pr[W_2]|
\end{equation}

\noindent from where the upper bound
\vspace*{2mm} 
\begin{equation}  \label{bounds:1:1}
PRGadv[\mathcal{A},G]\leq |Pr[W_0]-Pr[W_1]|+|Pr[W_1]-Pr[W_2]|
\end{equation}

\noindent follows easily.

Let $\mathcal{A}$ be a polynomial-time adversary to $G$, and $\mathcal{B}$
be a procedure that acts as a challenger to $\mathcal{A}$ and as an
adversary to RG. It follows the sequence given below:

\begin{itemize}
\item Receives $x_0$ from $\mathcal{A}$.

\item Runs \texttt{Algorithm \ref{algo1}} from lines \texttt{9} to
\texttt{14}.

\item Queries the RG challenger using $x$ as seed.

\item Upon receiving $z^{\prime }$ from the challenger, generates $%
z\leftarrow z^{\prime }$ and continues with \texttt{Algorithm \ref{algo1}} from line
\texttt{15}.

\item Let $y_1,\ldots,y_m$ be the result. Send $y_1,\ldots,y_m$ to $\mathcal{%
A}$.

\item Output whatever $\mathcal{A}$ outputs.
\end{itemize}

\noindent In this scenario, it can be seen that
\vspace*{2mm} 
\begin{equation}  \label{bounds:1:2}
PRGadv[\mathcal{B},RG]=|Pr[W_0]-Pr[W_1]|
\end{equation}

Similarly, let $\mathcal{B}^{\prime }$ be an adversary to the PRF $f$
acting as a challenger to $\mathcal{A}$. Now $\mathcal{B}^{\prime }$ runs
the following sequence of steps:

\begin{itemize}
\item Receives $x_0$ from $\mathcal{A}$.

\item Runs \texttt{Algorithm \ref{algo1}} from lines \texttt{9} to
\texttt{16}.

\item Makes $m$ queries to the PRF challenger with $z_1,\ldots,z_m$.

\item Upon receiving $y_1,\ldots,y_m$ from the challenger, send $%
y_1,\ldots,y_m$ to $\mathcal{A}$.

\item Output whatever $\mathcal{A}$ outputs.
\end{itemize}

\noindent As a result, we have that
\vspace*{2mm} 
\begin{equation}
PRFadv[\mathcal{B}^{\prime },f]=|Pr[W_1]-Pr[W_2]|
\end{equation}

\noindent and since $\mathcal{B}^{\prime }$ cannot control the
sequence $z_1,\ldots,z_m$ sent to the challenger, the above equation can be
made even tighter. In fact,
\vspace*{2mm} 
\begin{equation}  \label{bounds:1:3}
PRFadv_{weak}[\mathcal{B}^{\prime },f]=|Pr[W_1]-Pr[W_2]|.
\end{equation}

\noindent The proof then follows by (\ref{bounds:1:1}), (\ref{bounds:1:2})
and (\ref{bounds:1:3}).
\end{proof}

\subsection{Stretch bounds}

In addition to the above security bound, we will also prove here that our
generator can extract more bits than the exponential generator given in \cite%
{Gennaro} and still remain secure, provided that some conditions on the
Feistel-like function hold.

Suppose that we stretch the output by one word, that is, in line
\texttt{\ref{algo1:14}} of \texttt{Algorithm \ref{algo1}} we keep the $t+k$ least
significant bits. Let $G^*$ be that generator. Our main result is that the
stretch keeps the generator safe provided that $f$ is a secure PRF even when
the adversary has access to an oracle for computing any member of $f$ at
random $m=t/k$ points (save, of course, the challenge value).

To keep the notation clear, let us write as $PRFadv^*[\mathcal{A},f]$ the
PRF advantage that an adversary $\mathcal{A}$ has over $f$ when having
access to an oracle for $f$ that allows it to make a (polynomially bounded)
number of queries, and postulate the following:

\begin{theorem}
If the RG PRBG (pseudorandom bit generator) is secure and $f$ is a secure
PRF, even in the existence of an oracle for $f$ the PRBG given in
\texttt{Algorithm \ref{algo1}}, where $t$ is replaced by $t+k$, is secure.
\end{theorem}

\noindent \texttt{Important remark} $-$ It is not difficult to note 
that for any polynomial-time adversary $\mathcal{A}$ able to break the PRBG,
there are also polynomial-time adversaries that are elementary wrappers
around $\mathcal{A}$, namely: an adversary $\mathcal{B}$ to the RG; an
adversary $\mathcal{B}^{\prime}$ to the PRF $f$ that has access to an oracle
for $f$ with at most $m=t/k$ queries; and an adversary $\mathcal{B}^{\prime
\prime }$ to the PRF $f$ that makes at most $m+1$ queries, such that 
\vspace*{2mm} 
\begin{equation*}
PRGadv[\mathcal{A},G^*] \leq PRFadv^*[\mathcal{B}^{\prime },f] + PRFadv[%
\mathcal{B}^{\prime \prime },f] + PRGadv[\mathcal{B},RG]
\end{equation*}

\noindent As with the previous Theorem, we resort to a sequence of games to
provide a formal proof.

\begin{proof}
Let $Game\ 0$ be the following interaction:

\begin{itemize}
\item The challenger runs one round of the extended generator (lines \texttt{9} to
\texttt{21} of \texttt{Algorithm \ref{algo1}} with $t$ replaced by $t+k$) with $%
y_1,\ldots,y_m,y_{m+1}$ as the result.

\item The challenger sends $y_1,\ldots,y_m,y_{m+1}$ to $\mathcal{A}$.

\item $\mathcal{A}$ outputs one bit $b$.
\end{itemize}

\noindent Define $Game\ 1$ as follows:

\begin{itemize}
\item The challenger sets $r$ as a random element of $\{0,1\}^k$.

\item The challenger runs one round of the extended generator (lines \texttt{9}
to \texttt{21} of \texttt{Algorithm \ref{algo1}} with $t$ replaced by $t+k$) and in
line \texttt{17} replaces $z_{m+1}$ with $r$, then continues with the algorithm. Let $%
y_1,\ldots,y_m,y_{m+1}$ be the result.

\item The challenger sends $y_1,\ldots,y_m,y_{m+1}$ to $\mathcal{A}$.

\item $\mathcal{A}$ outputs one bit $b$.
\end{itemize}

\noindent $Game\ 2$ evolves according to the following steps:

\begin{itemize}
\item The challenger sets $r_1,\ldots,r_m,r_{m+1}$ as random elements of $%
\{0,1\}^k$.

\item The challenger runs one round of the extended generator (lines \texttt{9}
to \texttt{21} of \texttt{Algorithm \ref{algo1}} with $t$ replaced by $t+k$) but
it now replaces $z_i$ with $r_i$ in line \texttt{16}, for $i=1,\ldots,m+1$. Let $y_1,%
\ldots,y_m,y_{m+1}$ be the result.

\item The challenger sends $y_1,\ldots,y_m,y_{m+1}$ to $\mathcal{A}$.

\item $\mathcal{A}$ outputs one bit $b$.
\end{itemize}

\noindent Finally, $Game\ 3$ proceeds as outlined below:

\begin{itemize}
\item The challenger sets $r_1,\ldots,r_m,r_{m+1}$ as random elements of $%
\{0,1\}^k$.

\item The challenger runs one round of the extended generator (lines \texttt{9}
to \texttt{21} of \texttt{Algorithm \ref{algo1}} with $t$ replaced by $t+k$). Let $%
y_1,\ldots,y_m,y_{m+1}$ be the result.

\item The challenger sends $r_1,\ldots,r_m,r_{m+1}$ to $\mathcal{A}$ (it
replaces $y_{i}$ with $r_i$).

\item $\mathcal{A}$ outputs one bit $b$.
\end{itemize}

Let $W_i$ be the event $\mathcal{A}$ outputs $1$ in $Game\ i$. Clearly, the
advantage of $\mathcal{A}$ in breaking the PRG $G^*$ is given by
\vspace*{2mm} 
\begin{equation*}
PRGadv[\mathcal{A},G^*]=|Pr[W_0]-Pr[W_3]|
\end{equation*}

\noindent Note that it is possible to express the right hand side of the
above expression as
\vspace*{2mm} 
\begin{equation*}
PRGadv[\mathcal{A},G^*]=|Pr[W_0] -Pr[W_1] + Pr[W_1] - Pr[W_2] + Pr[W_2] -
Pr[W_3]|
\end{equation*}

\noindent with bounds that can be easily derived from the triangle
inequality
\vspace*{2mm} 
\begin{align*}
PRGadv[\mathcal{A},G^*]& \leq |Pr[W_0] -Pr[W_1]| \\
& + |Pr[W_1] - Pr[W_2]| \\
& + |Pr[W_2] - Pr[W_3]|
\end{align*}

For readability, we label each term above as
\vspace*{2mm} 
\begin{align}
& |Pr[W_0] -Pr[W_1]|  \label{eq:bounds:1} \\
& |Pr[W_1] - Pr[W_2]|  \label{eq:bounds:2} \\
& |Pr[W_2] - Pr[W_3]| .  \label{eq:bounds:3}
\end{align}

\noindent and to begin with, let us provide an upper bound for (\ref%
{eq:bounds:1}). Assume $\mathcal{B}^{\prime }$ is an adversary to the PRF
used in step \texttt{17} of \texttt{Algorithm \ref{algo1}}. $\mathcal{B}^{\prime }$ acts
as a challenger to $\mathcal{A}$ by running the following sequence:

\begin{itemize}
\item Receives $x_0$ from $\mathcal{A}$.

\item Runs \texttt{Algorithm \ref{algo1}} up to line \texttt{16} to obtain $%
z_1,\ldots,z_m,z_{m+1}$.

\item Sends $z_1,\ldots,z_m$ as $m$ queries to its PRF oracle $\mathcal{O}$
and let $y_1,\ldots,y_m$ be the response of $\mathcal{O}$.

\item Sends $z_{m+1}$ as a query to its (PRF) challenger.

\item Upon receiving $y_{m+1}$ from the challenger, let $\bar{y}%
=y_1,\ldots,y_m,y_{m+1}$. Then, sends $\bar{y}$ to $\mathcal{A}$.

\item Output whatever $\mathcal{A}$ outputs.
\end{itemize}

\noindent As a result, we have that
\vspace*{2mm} 
\begin{equation}  \label{bound:1}
PRFadv[\mathcal{B}^{\prime },f]=|Pr[W_0]-Pr[W_1]|
\end{equation}

\noindent The second part of the inequality (\ref{eq:bounds:2}) can be
bounded by the advantage that an adversary gains over the RG PRBG.

Let $\mathcal{B}$ be an adversary to the RG PRBG that, in turn, acts as a
challenger to $\mathcal{A}$. $\mathcal{B}$ runs the following sequence:

\begin{itemize}
\item Receive $x_0$ from $\mathcal{A}$.

\item Set $r$ as a random element of $\{0,1\}^k$.

\item Run \texttt{Algorithm \ref{algo1}} until line \texttt{14}.

\item Query to its (PRG) challenger with $x$ as seed.

\item Upon receiving $z^{\prime }=z^{\prime }_1,\ldots,z^{\prime }_m$ from
the challenger, generate $z\leftarrow z^{\prime }$ and continue the
execution of \texttt{Algorithm \ref{algo1}} from line \texttt{16}.
Let $y_1,\ldots,y_m$ be the output.

\item Send $y_1,\ldots,y_m,r$ to $\mathcal{A}$ and output whatever $\mathcal{%
A}$ outputs.
\end{itemize}

\vspace*{1mm} \noindent It is therefore clear that
\vspace*{2mm} 
\begin{equation}  \label{bound:2}
PRGadv[\mathcal{B},RG]=|Pr[W_1]-Pr[W_2]|
\end{equation}

To conclude, it is easy to build an adversary $\mathcal{B}^{\prime \prime }$
to the PRF $f$ that makes at most $m+1$ queries to the PRF challenger and
uses $\mathcal{A}$ as a subroutine, in such a way that
\vspace*{2mm} 
\begin{equation}  \label{bound:3}
PRGadv[\mathcal{B}^{\prime \prime },f]=|Pr[W_2]-Pr[W_3]|
\end{equation}

\noindent We leave those details to the reader. The main result follows from
the equalities in (\ref{bound:1})-(\ref{bound:3}).
\end{proof}

\section{Tests}

We turn now to the statistical testing of sequences obtained using the
CSPRNG described in \texttt{Algorithm \ref{algo1}} and \texttt{Algorithm \ref{algo2}}. In our
initial discussion we argued that statistical tests, if anything, provide a
first line of defence against non-randomness (in other words, as a means to
discard bad generators) but only that. Let us now try to be more explicit.

Statistical testing comes with several questions attached. For example, is
there a necessary criteria to judge randomness? Not quite, for otherwise
only one battery would be enough, rather than competing sets. Despite some
procedures being included in all the testing suites outlined in the
introduction, such commonality does not seemingly exhaust the set of
necessary tests; they would provide, if anything, a \textit{minimal} set.
But also, is such criteria to be considered an absolute metric or one
relative to the application at hand? Should it be the latter, how do we go
about ranking different generators? (It would also require some method for
ranking every possible application of a PRNG, a nearly impossible task.)
Speaking of a metric, is there an agreed or available unit for measuring the
adequacy (or quality, depending on the approach) of a PRNG?

The above questions are an example of the difficulties a practitioner faces
when evaluating the statistical qualities of any PRNG. Our approach has been
to look at two test batteries: the NIST suite by \cite{NISTtest} and TestU01 by 
\cite{TestU01}. For further details the reader is referred to the information
and other materials available through the links outlined in the
introduction. The reason for selecting those batteries was a practical one:
concerning the NIST test, originally developed in the '90s, is the approved
testing protocol for certification under NIST standards; for that very
reason, it has become the starting point for other testing suites and hence
provides an obvious benchmark. TestU01, first developed in 2007, goes futher
and deeper than the NIST suite, and is nowadays acknowledged as providing
the most reliable testing battery for random sequences.

Going back to the questions about the statistical properties of any PRNG,
they ought to display some basic (or minimal, as stated above) properties.
Those are the following: \vspace*{-0.8mm}

\begin{itemize}
\item The symbols generated should be, quite obviously, independent of each
other. It is equivalent to say that there should not be serial correlation
between successive symbols in any generated sequence.

\item The frequency by which a PRNG generates any symbol\ should not be
higher than for any other symbol\ in the defined or assigned output range.
It is equivalent to requiring those numbers to be equiprobable.

\item A further property is that of uniformity in the distribution of the
generated symbols, meaning they should be evenly spread (or symmetric).

\item In addition, any permutations of the symbols generated by a PRNG
should also be equiprobable (as otherwise the generator would be biased).

\item The period of the PRNG, or how long it takes the generator to produce
a sequence identical to the first one it created, should be large. The
question becomes qualifying what \textit{large} means. In this regard, refer
to the opening paragraph in sub-section 3.1 for the calculation of our
generator complexity. The periodicity of our PRNG as tested is just above $%
2^{158}$. Bear in mind, however, that periodicity does not equate to
security. Once the period exceeds $2^{64}$ or $2^{128}$ for a 64-bit
generator\ it becomes completely irrelevant for cryptographic applications.

\vspace*{1mm} \noindent It should be noted here that any deterministic
machine, such as a computer, has a number of states bound by its finite
memory (as opposed to Turing Machines that have infinite memory; they still
have a finite number of states, however they are arbitrarily large).
Consequently, any program running on a computer will eventually return to a
state where it has been before. How long would it take for that to happen
(or more precisely, the number of computations required) is key to determine
predicatibilty: if repeating itself takes an impossibly long time for a
PRNG, any sequences it generates would not be predictable using available
(or even future) computing power.

\item To start a PRNG a seed is required. From a statistical perspective,
the above listed properties should hold independently of the seed(s) used to
initialize a PRNG.
\end{itemize}

Statistical testing is used to determine if a sequence, either defined over
the real interval $\left( 0,1\right) $ or over the binary set $\left\{
0,1\right\} $, significantly departs from a true random realization, and it
involves probabilities. Both the NIST and TestU01 batteries look at the
presence of different types of patterns in sequences generated by PRNG; if
any of those patterns are detected, then the sequence is considered not to
be random.

The assumption that a sequence generated by a PRNG is indeed random becomes
the null hypothesis, denoted as $H_{0}$. Of course, given $H_{0}$ one faces
two scenarios: firstly, that a truly random sequence is deemed not to be
random (rejecting $H_{0}$ when it should have been accepted) leading to a
Type I error, or a false positive; secondly, that a non-random sequence is
accepted as random (accepting $H_{0}$ when the opposite is true) defined as
a Type II error, or a false negative. Type I errors are controlled by
setting the significance level $\alpha $ of the tests. Since $H_{0}$ is
evaluated in terms of probabilities, the strength of the evidence provided
by the data against $H_{0}$ is given by the $P_{value}\in \left[ 0,1\right] $
of the test. If $P_{value}\geq \alpha $ then $H_{0}$ is accepted; given our
null, it implies that the sequence being tested is random with a confidence
level equal to $1-\alpha $.

Should the value of $\alpha $ be set too high, a data sequence that is truly
random faces the possibility of being rejected as such, incurring Type I
errors; if set too low, one might end accepting as random a data sequence
that is not so, hence inducing a bias towards Type II errors. Furthermore,
due to the existence of many potential sources of non-randomness, in this
particular testing scenario Type II errors are more problematic to appraise
than Type I errors.

To avoid accepting as a good generator one that is flawed it is advisable to
use additional testing procedures, running the $t=1,2,...,$ $T$ tests in a
battery over each sub-sequence, and counting the number of times $%
P_{value,t}\geq \alpha ,\forall t$. Using the ratio $R_{t}=\left( 1/M\right)
\sum_{M}\left( P_{value,t}\geq \alpha \right) $ a test $t$ is considered
passed if $R_{t}$ is greater than $1-\alpha $. In addition, and as noted by 
\cite{duncan}, under $H_{0}$ the $P_{value,t}$ statistics across the ${M}$
sub-sequences should be uniformly distributed over the $\left( 0,1\right] $
interval, so a simple $\chi ^{2}$ test or other non-parametric options (for
example, the Kolmogorov-Smirnov test) can be used in this instance to check
for bias.

It is also possible, although in many instances\ computationally expensive,
to test $h$ runs of size $m$ using different seeds, or changing other
parameters.

This meta-approach is equivalent to looking at the results from $h\left(
MT\right) $ tests; should testing large $m$ sequences be possible, adding
the $h$ dimension makes it easier to more properly look at each test on
their own.

\vspace*{2mm} \noindent 5.1. \textbf{Materials and methods.} We defined as 
objective to evaluate at least $h=100$ runs of our generator,
on sequences of about $m=2^{40}$ bits in size 
(or $2^{35}$ 32-bit integers, hence $2^{35}\times 32=2^{40}$ bits. Each
sequence required storing just over 137 Gb\ of data). The algorithm picked
skips $\left\{ w_{1},w_{2}\right\} $ starting from a prime $p\in \mathcal{P}$
of size on the interval $k=\left[ 32,64\right]$ to ensure they are evenly
distributed when taken as 32-bit strings; please refer to Section 3 for
details (note that if $p<32$ bits and the $w_{k}$ skips are taken as 32-bit
strings, it would always be the case that bit $32$ is a zero, resulting in
bias).

One of the features of our PRNG design is its reasonable speed. Given the
size of the sequences to be tested and to ensure both the NIST and TestU01
batteries took as little time as possible to complete, our initial focus
became selecting a $q_{size}$ (size in bits of the safe primes) that
reasonably ensured our PRNG produced sequences showing good statistical
qualities, even if such choice could render an attack on the PRNG feasible
The table below shows the results obtained for sequences of $2^{33}$ bits
against several $q_{size}$ options.

To test execution times, we took the first safe prime $q$ of the form $2p-1$
with $p$ denoting a prime of $u-1$ bits, $u=\left\{
256,512,1024,2048\right\} $. The average execution time includes, give or
take, a $30\%$ overhead due to disk writes. Table compiled from runs made on
an Intel Core i5-8400 running at 2.80 GHz with 16 Mb RAM.

\vspace*{1mm}

\begin{center}
\begin{tabular}{ccc}
Safe prime $q_{size}$ &  & Execution time \\ 
{\lbrack}in bits{\rbrack} &  & {\lbrack}in seconds{\rbrack} \\ \hline
$256$ &  & $296$ \\ 
$512$ &  & $343$ \\ 
$1024$ &  & $410$ \\ 
$2048$ &  & $538$ \\ \hline
\end{tabular}

\makecell{\texttt{Table I:} $q_{size}$ \texttt{against PRNG execution time}}
\end{center}

It soon became apparent that safe primes of $q_{size}\approx 256$ bits
already provided robust and consistent outcomes, for speed gains of roughly
four hours per run compared to using (cryptographically safe)\ primes of at
least $q_{size}=$ $1024$. Therefore, we adopted a $q_{size}=$ $260$ bits
running the experiments in a cluster of $12$ PCs.

Seven PCs being Intel Core i5-8400 CPU running at 2.80GHz and the remaining
four of them Intel Core i5-4590 CPU running at 3.30GHz, all with 16 Mb of
RAM. One spare PC was used for re-runs. TestU01 using the Crush battery
added about an hour of execution time, including overheads due to disk
writes.

To run TestU01, our algorithm was coded in C++, and compiled (together with
the TestU01 routines and libraries) using gcc 7.4.0 in Ubuntu 18.04.1. A
more versatile Python 3 version of the algorithm has also been coded.

The NIST battery, however, demanded a markedly longer time to complete,
hence we only ran $10$ experiments simply as confirmatory analysis, using
five sequences already tested using TestU01 and five new ones. Note the NIST
suite is typically used for (and seems prepared to handle) sequences of size 
$2^{20}$ or only slightly longer. Concerning execution times, that might
explain its weaker performance given our testing scenario.

\vspace*{1mm}

\begin{center}
\begin{tabular}{ccccccc}
\multicolumn{3}{c}{\textbf{Cluster A}} &  & \multicolumn{3}{c}{\textbf{%
Cluster B}} \\ \cline{1-3}\cline{5-7}
PC Id & Runs & Weak &  & PC Id & Runs & Weak \\ \cline{1-3}\cline{5-7}
$1$ & $12$ & $0$ &  & $17$ & $3$ & $1$ \\ 
$2$ & $15$ & $0$ &  & $21$ & $5$ & $0$ \\ 
$3$ & $15$ & $0$ &  & $22$ & $5$ & $0$ \\ 
$4$ & $15$ & $0$ &  & $23$ & $5$ & $1$ \\ 
$5$ & $15$ & $0$ &  & \multicolumn{3}{c}{\textbf{Backup PC}} \\ \cline{5-7}
$14$ & $9$ & $2$ &  & PC Id & Runs & Weak \\ 
$15$ & $3$ & $0$ &  & $14$ & $3$ & $0$ \\ \cline{1-3}\cline{5-7}
\end{tabular}

\makecell{\texttt{Table II: TestU01 results for each PC} \\\texttt{in our
testing clusters}}
\end{center}

Concerning the test batteries themselves, the NIST\ suite is currently
composed of 15 statistical tests that look at multiple potential sources of
non-randomness in (arbitrarily long) binary sequences. In its 2010 revision,
the NIST removed the \textit{Lempel-Ziv Complexity of Sequences} (\#10 in
the battery) due to a detected bias in the $P_{value}$ of the test.

The Crush battery of TestU01 includes 96 separate tests, and provides a good
template to judge if a PRNG is broken; in fact, the gap between this battery
and BigCrush (160 tests) is rather small, in practical terms, if compared to
the gap between the Crush battery and competing testing suites, including
the NIST one. It is much more likely to find a PRNG that, having passed
other testing protocols convincingly, fails in at least one of the tests of
the Crush battery, than a PRNG failing a BigCrush test having otherwise
passed under Crush.

\newpage
\vspace*{2mm} \noindent 5.2. \textbf{Results.} The outcome of TestU01 is summarized 
in \texttt{Table II}.\ For $h^{\ast
}=105$ total runs using sequences of about $m=2^{40}$ bits in size, $101$
runs showed no problems and recorded four suspicious results at a
significance level of $\alpha =0.001$ involving the tests in \texttt{Table
III} from TestU01 Crush battery. Note each $P_{value}$ shows those
suspicious results have been marginal (a weakness, rather than an outright
failure)\ at the set significance level, in addition of not being systematic.%
An outright failure implies the $P_{value}$ is outside the $\left[
10^{-10},1-10^{-10}\right] $ range (meaning too close to either 0 or 1). In
our case, the $P_{value}$ of the tests is quite near or at the set
significance level. The fact they were also isolated results, rather than a
repeated outcome, also plays a role in the evaluation.

\vspace*{1mm}

\begin{center}
\begin{tabular}{cccc}
Cluster & PC\ Id & Test (\# and descriptor) & $P_{value}$ \\ \hline
A & $14$ & \multicolumn{1}{l}{\textbf{\#92} \texttt{sstring\_Run}} & 
\multicolumn{1}{r}{$5.0e-4$} \\ 
A & $14$ & \multicolumn{1}{l}{\textbf{\#94} \texttt{sstring\_AutoCor}} & 
\multicolumn{1}{r}{$0.9990$} \\ 
B & $17$ & \multicolumn{1}{l}{\textbf{\#84} \texttt{sstring\_HammingCorr}} & 
\multicolumn{1}{r}{$0.9991$} \\ 
B & $23$ & \multicolumn{1}{l}{\textbf{\#80} \texttt{sstring\_HammingWeight2}}
& \multicolumn{1}{r}{$3.8e-4$} \\ \hline
\end{tabular}

\makecell{\texttt{Table III: TestU01 suspicious results}}
\end{center}

\noindent Proceeding in the same order as shown above, the description of
the tests is the following (the \# identifier for the tests is that in \cite{TestU01}, pp.
144-147)

\vspace*{1mm}

\begin{itemize}
\item \textbf{\#92} is simply a version of the runs tests applicable to bit
strings.

\item \textbf{\#94} checks the autocorrelation between bits of order $d$.

\item \textbf{\#84} also applies a correlation test, but based on the
Hamming weights of successive blocks of $L$ bits.

\item \textbf{\#80}, finally, examines the proportion of 1's within
non-overlapping blocks of $L$ bits.
\end{itemize}

\noindent To fully ascertain the situation, we conducted $10$ repeats for
each sequence (complete re-runs) using, on every occasion, identical
parameterizations to the ones that led to suspicious outcomes: five runs on
the same suspecting PCs, the other five repeats on any of the remaining PCs
picked at random. No problems were detected in any of the $10$ repeat tests.
Further investigation, however, revealed that micro-cuts in the power supply
occurred during the same time intervals the tests in Table \texttt{III} were
running; this finding is a provable, but not proven, explanation for the
suspicious results, as our inability to subsequently replicate them seem to
suggest.

Therefore, upon considering (a) the inability to mimic those results in $10$
repeat runs; (b) the detected issues are not systematic; and (c) a possible
external cause linked to those outcomes, it seems appropriate to assume the
four tests as technically\ passed.

Turning now to the successes, the $P_{value}$ for every test in the Crush
battery positioned itself comfortably in the acceptance region, for the vast
majority of cases. As a last step, we computed Kolmogorov-Smirnov and
Anderson-Darling statistics across the $h^{\ast }=105$ runs, assuming, under
the null hypothesis (and as previously noted) the $P_{value}$ for each of
the tests is uniformly distributed. In all instances, that $H_{0}$ was
decisively accepted at a $\alpha =0.01$ significance.

Concerning the NIST battery, the tests were run on a single PC, showing no
fails and one weakness in the $10$ runs processed. As before, the suspicious
result (involving a single, isolated instance of the Non-overlapping
Template Matching test) was very marginal, happening only once across the $%
10 $ runs and hence bearing no practical importance.

\section{Conclusions}

The focus of this paper has been to present a novel algorithm for the
generation of pseudorandom sequences that is (a) statistically sound; and
(b)\ provides a high degree of cryptographic security. Concerning the
former, we have relied on the results from the NIST (on a very limited
scale)\ and the TestU01 Crush battery of tests; TestU01 includes all the
procedures in the NIST\ suite plus supplementary tests in its Crush battery.
On the latter, the discussion of Section 3 (and in particular sub-section
3.1) provides theoretical support to claim our generator is crytographically
secure. In that sense, we have found no cryptoanalytic technique able to
break the particular combination of modular exponentiations and modified
Feistel boxes in \texttt{Algorithm \ref{algo1}} and \texttt{Algorithm \ref{algo2}}, leaving
Brute Force as the only plausible attack vector.

This possibility is also denied, however, given the computing power nowadays
available and most likely to become available for quite a long time.

Inevitably, in any CSPRNG design there is a balance to strike between speed
and suitability for cryptographic use. Given the present code
implementation, our algorithm has shown an average speed of 1063 cycles/byte
to get 32-bit random numbers in non-dedicated hardware using 1024-bit safe
primes. Such performance might seem comparatively poor, at first sight,
compared to other options available; however, the statistical tests
performed on sequences of about $2^{40}$ bits length have shown no evidence
of issues other than the reported flukes.

Our proposed design therefore compares very well against most publicly known
PRNGs, as they invariably display between one to seven weak results (and the
occasional failure) using the same test batteries and sequence length as in
our experiments; in fact, popular PRNG generators have shown multiple
failures in TestU01 tests using the Crush battery. Of course, the
performance of our generator remains to be seen for sequences in the order
of several Tb in size, or perhaps using a more stringent testing battery
(e.g. BigCrush). That exercise is left for future work.

As a bonus, our design is easily portable. We have coded the \texttt{Algorithm \ref{algo1}} and 
\texttt{Algorithm \ref{algo2}} (together with supporting libraries)
in Python 3 and C++, in the latter case to hardwire our generator in TestU01.

Wrapping up, our expectation is that the scientific community could benefit
from a robust, portable and reasonably quick PRNG for most practical
applications, having, at the same time, theoretical cryptographic security
guarantees.

\section{Appendix}

The reduction is the procedure outlined below.

\begin{enumerate}
\item $x_0 \leftarrow x\cdot t$. \vspace*{1mm}

\item Let $x_1=g^{tx}$ be the output of the generator for the seed $x_0$
(note that the first bit of $x_0$ is $0$ and that $x_0\ div \ t=x$). 
\vspace*{1mm}

\item $x_2\leftarrow x_1$. Repeat the following $l$ times: \vspace*{1mm}

\begin{enumerate}
\item Let $w_1,w_2$ be the square roots of $x_2$ in $\mathbb{Z}_p$ \vspace*{%
1mm}

\item $x_{2}\leftarrow \Big\{\vspace*{1mm} 
\begin{array}{ll}
w_{1} & \text{\textnormal{if }}w_{1}\text{\textnormal{\ is a square in }}%
\mathbb{Z}_{p}. \\ 
w_{2} & \text{\textnormal{otherwise}}%
\end{array}%
$
\end{enumerate}

\item Return $x_2$.
\end{enumerate}

All the operations involved above are polynomially bounded, since $p$ is not
just a prime but a strong prime, so $p\equiv 3 \mod 4$ and the square roots
of an element in $\mathbb{Z}_p$, if they exist, are easy to find and there
are exactly two (see \cite{Bach} Corollary 7.1.2). Finally, for the same
reason at most one of the roots is at the same time a square in $\mathbb{Z}%
_p $ because $-1$ is a quadratic non-residue of $p$ (see \cite{Bach},
Theorem 5.8.1).

\end{document}